\documentclass[pra,twocolumn,epsfig,rotate,superscriptaddress,showpacs]{revtex4}
\usepackage{graphicx}
\usepackage{epstopdf}
\usepackage{amsfonts}
\usepackage{amssymb}
\usepackage{amsmath}
\usepackage{subfigure}
\usepackage{prettyref}
\usepackage{float}
\usepackage{color,CJK}
\usepackage{amsmath}
\usepackage{amssymb}
\usepackage{esint}

\usepackage{braket}

\def\be{\begin{equation}}
\def\ee{\end{equation}}
\def\ba{\begin{eqnarray}}
\def\ea{\end{eqnarray}}

\begin{document}
\begin{CJK}{UTF8}{gbsn}

\title{Majorana modes in solid state systems and its dynamics}

\author{Qi Zhang(张起)}
\affiliation{College of Science, Zhejiang University of Technology,
Hangzhou 310014, China}

\author{Biao Wu(吴飙)} 
      \affiliation{International Center for Quantum Materials, School of Physics, Peking University, Beijing 100871, China}
      \affiliation{Collaborative Innovation Center of Quantum Matter, Beijing 100871, China}
      \affiliation{Wilczek Quantum Center, School of Physics and Astronomy, Shanghai Jiao Tong University, Shanghai 200240, China}

\date{\today}
\begin{abstract}
We review the properties of Majorana fermions in particle physics and
point out that Majorana modes in solid state systems are significantly different.
The key reason is the concept of anti-particle in solid state systems is different
from its counterpart in particle physics.  We define  Majorana modes
as the eigenstates of Majorana operators and find that they can exist both at edges and in the bulk.
According to our definition, only one single Majorana mode can exist in a system no matter
at edges or in the bulk. Kitaev's spinless $p$-wave superconductor is used to illustrate our
results and the dynamical behavior of the Majorana modes.
\end{abstract}

\maketitle
\end{CJK}
\section{Introduction}
As an elementary particle, Majorana fermion is a fermion whose anti-particle is itself.
Mathematically, it is the result of quantizing a real Dirac field. This is similar to the case
where quantizing the real electromagnetic field leads to photon whose anti-particle
is itself. So far, there is no direct observation of  Majorana fermion as an elementary
particle~\cite{MF}.  In recent years, there have been intensive efforts  to explore
the possible existence of Majorana fermion-like state as edge modes in condensed
matter systems~\cite{Kitaev,OMF1,OMF2,OMF3,OMF4,OMF5,OMF6,OMF7,OMF8,xjliu1,xjliu2,chen,HuYing,Maiti}.
Observation of the Majorana zero modes has been reported in many experiments
~\cite{Exp1,Exp2,Exp3,Exp4,Exp5,Jia}.

In this work we review the basic features of Majorana fermion as an elementary particle.
We point out that due to the lack of the existence of {\it real} antiparticle,
the Majorana fermions in condensed matter systems are different from Majorana fermions in high
energy physics although they do share some common features.  In this perspective, to avoid
confusion, it is better to call Majorana fermion-like states found in condensed matter systems
Majorana modes.

We define a Majorana mode as an eigenstate of Majorana operator
$\gamma^\dagger=c^\dagger+c$, where $c^\dagger$ and $c$ are creation and annihilation operators
of electrons, respectively. Our definition of Majorana mode is illustrated with Kitaev's model of one dimensional
spinless $p$-wave superconductor~\cite{Kitaev}. Usually researchers mainly focus on the Majorana edge modes as topologically protected states and whether robust against modulation or disorders~\cite{chen,HuYing}. It is clear from our definition that the Majorana modes can exist both at the edges and in the bulk. At the edge, our Majorana mode is an
energy eigenstate. In the bulk, our Majorana mode is not an energy eigenstate thus can
evolve dynamically. The dynamics of our Majorana mode in the bulk is characterized by
hopping back and forth along the chain accompanied by the Majorana mode
switching its type cyclically.

Because an electron can be mathematically written as a pair of Majorana operators, it is a
common belief that Majorana fermions only occur in pairs. For instance, in the Kitaev chain system, it is generally believed that there is a pair of Majorana fermions (or modes) of different types
at both edges of the chain. However, according to  our definition,  we can show that
only one single Majorana mode can exist along the whole chain. A Majorana mode in the bulk excludes the existence of that on the edge and vice versa.

\section{Majorana fermion in particle physics}
We start by reviewing the basic concepts of particle and anti-particle in particle physics,
which can be found in any standard book on quantum field theory~\cite{Zee}.
Particle and anti-particle arise naturally when we
quantize a free Dirac field $\psi(x)$. For clarity,  we consider only the one dimensional case
and omit time variable.  The quantization starts with a Fourier expansion ($\hbar=c=1$)
\be
\psi(x)=\sum_{k,r=1,2} \frac{1}{\sqrt{2VE_k}}[b_r(k)u_r(k)e^{-ikx}+d_r^\dag(k)v_r(k)e^{ikx}],
\ee
where $V$ is the volume, $E_k$ is the energy, and $k$ is the momentum.  $u_r(k)$ ($v_r(k)$) is the four-component spinor of the positive (negative) energy branch, and $b_r(k)$ ($d_r(k)$) is the field operator at a given momentum. Imposing the anti-commutation relation for fermions $\{b_r(k),b_{r'}^\dag(k')\}=\{d_r(k),d_{r'}^\dag(k')\}=\delta_{kk'}\delta_{rr'}$, we find that the energy and charge operator can be written as (neglecting an overall constant),
\ba  \label{Denergy}
E&=&\sum_{k,r=1,2}E_k[b_r^\dag(k)b_r(k)+d_r^\dag(k)d_r(k)], \\  \label{Dcharge}
Q&=&e\sum_{k,r=1,2}[b_r^\dag(k)b_r(k)-d_r^\dag(k)d_r(k)],
\ea
with $e$ being the elementary charge that cannot be fixed by Dirac equation.
It is apparent that, according to Eqs.~(\ref{Denergy}) and (\ref{Dcharge}), $b_r(k)$ and $d_r(k)$ are the operators for the Dirac fermion and its anti-fermion, respectively, which possess the same positive energy but the opposite charge.

For the purpose of comparison with the condensed matter system, it is important to note
two points: (1) particle and its anti-particle are physically two different entities, which have opposite charges.  (2) Mathematically, the creation operator for particle $b_r(k)^\dagger$ is not a hermitian conjugate of the annihilation operator of the anti-particle $d_r(k)$.

Majorana fermion arises when the Dirac field is real,  $\psi(x)=\psi^*(x)$.
In this case we have to write
\ba \nonumber
\psi(x)&=&\sum_{k,r=1,2} \frac{1}{\sqrt{2VE_k}}\Big[\frac{b_r(k)+d_r(k)}{\sqrt{2}}
u_r(k)e^{-ikx} \\
&+&\frac{b_r^\dag(k)+d_r^\dag(k)}{\sqrt{2}}u^*_r(k)e^{ikx}\Big]\,.
\ea
We let
\be
{\mathfrak m}^\dagger_r(k)=\big[b^\dagger_r(k)+d^\dagger_r(k)\big]/\sqrt{2}\,.
\label{mfh}
\ee
This is  the Majorana creation operator in particle physics.
This shows  that a Majorana fermion is effectively
a  superposition of an  fermion and its anti-particle.  That the Majorana fermion
is its own anti-particle is manifested in that ${\mathfrak m}^\dagger$ does not
change when $b^\dagger$ and $d^\dagger$ exchange their role under charge conjugate operation.
In other words, the Majorana fermion as an elementary particle is {\it not} manifested mathematically
with ${\mathfrak m}^\dagger={\mathfrak m}$. In particle physics, we always
have ${\mathfrak m}^\dagger\neq {\mathfrak m}$. It is very important to bear this in mind
when we discuss Majorana fermion in solid state physics.

\section{Majorana fermions in solid state systems}
It is clear from the above brief review that the existence of anti-particle is essential
to the concept of Majorana fermion. However, in solid state systems, the {\it real}
anti-particle of the electron, the positron, does not exist, and the positively charged {\it real}
particles are ions, which have different masses.
The closest thing to the anti-particle of electron is hole, which
emerges when an electron is annihilated from the Fermi sea, $c_k\ket{\rm Fermi~sea}$
(or $c_j\ket{\rm Fermi~sea}$). Physically, this hole is positively charged due to the existence
of positive ion background and has effectively the mass of an electron. This is similar
to the anti-particle in particle physics, which has opposite charge of the particle while sharing
the same mass. As indicated in Eq.(\ref{mfh}),  Majorana fermion in particle physics
is a superposition of a fermion and its anti-particle. If we follow
this rule,  a Majorana operator in solid state systems has to be defined as~\cite{Kitaev}
\be
\label{mf1}
\gamma^\dagger=c^\dagger+c\,.
\ee

However, physically and mathematically the Majorana fermion defined above is different
from Majorana fermion as an elementary particle. Mathematically,
we have $\gamma^\dagger=\gamma$ while ${\mathfrak m}^\dagger\neq {\mathfrak m}$.
Physically, the hole does not have its own identity: while we can create an electron with
$c^\dagger\ket{\rm{vaccum}}$, we get nothing with $c \ket{\rm{vaccum}}$. In contrast,
in particle physics, we can create simultaneously a particle and its anti-particle with
$b^\dagger\ket{\rm{vaccum}}$ and $d^\dagger\ket{\rm{vaccum}}$. In particle physics,
an anti-particle is a real particle existing on top of a vacuum with a given momentum or at a given position. In solid state systems, the hole state $c_k\ket{\rm Fermi~sea}$
(or $c_j\ket{\rm Fermi~sea}$) represents physically an empty (vacuum) state at a
given momentum $k$ (or a given physical site $j$).

There is another possibility of having Majorana fermions in solid state systems.
In superconductors, an electron can be screened by superconducting Cooper pairs to be
charge neutral~\cite{MF}. In this case, the electron effectively is its own anti-particle
and becomes a Majorana fermion. However, so far there is no quantitative calculation to
show such an efficient screening and no experiment demonstrates that the electron in
a superconductor behaves like a neutral particle. Still there is difference. In particle physics,
${\mathfrak m}^\dagger\ket{{\rm any~state}}$ and
${\mathfrak m}\ket{{\rm any~state}}$ are two different states. In a superconductor
where the charge screening is very effective, $c^\dagger\ket{\rm SC}$ and
$c\ket{\rm SC}$ are the same state as the superconducting state $\ket{\rm SC}$
is the same with one more or less Cooper pair~\cite{MF}.

There have been many studies~\cite{OMF1,OMF2,OMF3,OMF4,OMF5,OMF6,OMF7,OMF8,chen,xjliu1,xjliu2,HuYing,Maiti}, which show that the Majorana operator $\gamma^\dagger$ is related to many
interesting physics in solid state systems. Therefore, it is meaningful to study Majorana
fermion-related physics in solid state systems as long as we bear in mind all
the differences discussed above.

We focus on lattice models with $c^\dagger_j\,,c_j$ denoting creation and annihilation operators
at site $j$, respectively. Following Kitaev~\cite{Kitaev}, we introduce the following Majorana
operator
\begin{equation} \label{zhma}
\gamma_{j,\theta}=c_j^\dag e^{2i\theta}+c_j e^{-2i\theta}\,,
\end{equation}
which is more general than the one in Eq.(\ref{mf1}). One can easily check that
$\gamma_{j,\theta}=\gamma_{j,\theta}^\dagger$ and $\gamma_{j,\theta}^2=1$.
As a result, the Majorana operator $\gamma_{j,\theta}$ can have only two
eigenvalues $\pm 1$ and the corresponding eigenstates are
\begin{equation} \label{pm}
|j,\theta,\pm\rangle=|0\rangle_j e^{-i\theta}\pm|1\rangle_j e^{i\theta},
\end{equation}
where $|0\rangle_j$ denotes the vacuum at site $j$ and $|1\rangle_j$ denotes an electron at $j$.
We immediately notice that $|j,\theta,-\rangle$ is also the eigenstate of $\gamma_{j,\theta+\pi/2}$
with eigenvalue 1, i.e., $|j,\theta,-\rangle=|j,\theta+\pi/2,+\rangle$. This means that we can focus only on eigenstates $|j,\theta,+\rangle$.
For simplicity, we will always use  $|j,\theta\rangle=|j,\theta,+\rangle$ from now on.
If we introduce the following ``charge conjugate" operator
\begin{equation}
\mathcal{C} |1\rangle_j e^{i\theta}=|0\rangle_j e^{-i\theta},~~\mathcal{C} |0\rangle_j e^{-i\theta}
=|1\rangle_j e^{i\theta}\,.
\end{equation}
we apparently have $|j,\theta\rangle=\mathcal{C}|j,\theta\rangle$. This is the Majorana mode which
is the same as its anti-mode.

The Majorana-related physics in solid state systems is about the properties of these eigenstates
and how to create and manipulate them~\cite{OMF1,OMF2,OMF3,OMF4,OMF5,OMF6,OMF7,OMF8,chen,xjliu1,xjliu2,HuYing,Maiti}.  In the next section, we shall use
Kitaev's model~\cite{Kitaev}  to illustrate some of its properties. And we call
these eigenstates Majorana modes to avoid confusion with Majorana fermion
in particle physics.

\section{Kitave's superconducting chain}
In this section, we consider  the one dimensional Kitaev model, which describes a spinless
$p$-wave superconductor. Its Hamiltonian is given by~\cite{Kitaev,Flensberg}.
\begin{equation}
\label{Hamiltonian}
H=-\mu\sum_{j=1}^{L}c_{j}^\dag c_{j}-\sum_{j=1}^{L-1}
(t_p\ c_{j}^\dag c_{j+1}+\Delta\ c_{j}^\dag c_{j+1}^\dag+h.c.)\,,
\end{equation}
where $h.c.$ is for hermitian conjugate, $\mu$ is the chemical potential, $c_j$ is the electron
annihilation operator for site $j$, and $L$ is the length of the chain. The tunneling $t_p$ and superconducting gap $\Delta=|\Delta|e^{i\alpha}$ are the same for all the sites. For simplicity and without loss of generality, we assume $\Delta=|\Delta|$.
The Kitaev system can be realized experimentally by contacting a nanowire that has strong spin-orbit coupling (e.g., InSb and InAs nanowire) with a $s$-wave superconductor and in a Zeeman field~\cite{Exp1,Exp2,realization}. For clarity, in this work we focus on the condition $\mu=0$, $t_p=\Delta$~\cite{Kitaev,Flensberg}.

Mathematically, an electron can be written as a superposition of a pair of
Majorana operators,
\begin{equation} \label{Majorana}
c_j^\dag=\frac{1}{2}(\gamma_{j,0}^\dag-i\gamma_{j,\frac{\pi}{4}}^\dag)\,,
~~~c_j=\frac{1}{2}(\gamma_{j,0}^\dag+i\gamma_{j,\frac{\pi}{4}}^\dag)\,.
\end{equation}
where $\gamma_{j,0}=\gamma_{j,0}^\dag$ and
$\gamma_{j,\frac{\pi}{4}}=\gamma_{j,\frac{\pi}{4}}^\dag$ are defined in Eq.~(\ref{zhma}).
Majorana operators at neighboring sites can be combined to  form two new operators,
\begin{eqnarray}  \label{Ki-op} \nonumber
&&\tilde{c}_j^\dag =\frac{1}{2}(\gamma_{j,\frac{\pi}{4}}^\dag-i\gamma_{j+1,0}^\dag)\,,~~~
\tilde{c}_j =\frac{1}{2}(\gamma_{j,\frac{\pi}{4}}^\dag+i\gamma_{j+1,0}^\dag)\,; \\
&&\tilde{c}_L^\dag =\frac{1}{2}(\gamma_{L,\frac{\pi}{4}}^\dag-i\gamma_{1,0}^\dag)\,,~~~
\tilde{c}_L =\frac{1}{2}(\gamma_{L,\frac{\pi}{4}}^\dag+i\gamma_{1,0}^\dag)\,.
\end{eqnarray}
where $j=1,2,\cdots,L-1$. One can verify that  $\tilde{c}_j^\dag$ and $\tilde{c}_j$
are ordinary fermionic creation and annihilation operators, i.e., $[\tilde{c}_j,\tilde{c}_k^\dag]_+=\delta_{jk}$.
With these newly defined operators,
the Kitaev Hamiltonian in Eq. (\ref{Hamiltonian}) becomes,
\begin{equation} \label{Kitaev}
H=i\Delta \sum_{j=1}^{L-1}\gamma_{j,\frac{\pi}{4}}\gamma_{j+1,0}=2\Delta\sum_{j=1}^{L-1} \left(\tilde{c}_j^\dag\tilde{c}_j-\frac{1}{2}\right)\,.
\end{equation}
This shows that the energy eigenstates of this superconductor are
composed of integer number of quasi-particles denoted by $\tilde{c}_j^\dag$, $\tilde{c}_j$ instead of real electrons. The ground state $\ket{g}$ of this system satisfies
\begin{equation} \label{ground}
\tilde{c}_j|g\rangle=0, \quad {\text {for}} \ j=1,2,\ldots,L.
\end{equation}

Note that the Majorana operator $\gamma_{1,0}$ at the left end
and the Majorana operator $\gamma_{L,\frac{\pi}{4}}$  at the right end are missing in the
diagonalized Hamiltonian (\ref{Kitaev}).  As a result, the operators $\tilde{c}_L$ and
$\tilde{c}_L^\dagger$ are also missing.
This means that the Kitaev chain has two degenerate ground states, $\ket{g}$ and
$\tilde{c}_L^\dagger\ket{g}$.

\subsection{Majorana edge modes}

We now examine the properties of  the Majorana modes in the Kitaev chain.
Using the two degenerate ground states, we define two states
\begin{eqnarray}
|1,0\rangle&=&|g\rangle+i\tilde{c}_L^\dag|g\rangle\,, \\
|1,\frac{\pi}{2}\rangle&=&|g\rangle-i\tilde{c}_L^\dag|g\rangle\,.
\end{eqnarray}
These two states are orthonormal to each other. Furthermore, they are Majorana modes as
we have
\begin{eqnarray}
\gamma_{1,0}|1,0\rangle&=&|1,0\rangle\,,\\
\gamma_{1,\frac{\pi}{2}}|1,\frac{\pi}{2}\rangle&=&|1,\frac{\pi}{2}\rangle\,.
\end{eqnarray}
Interestingly, both of the two Majorana modes are at site 1, the left end of the chain. This is contrary
to the general belief that there is a pair of Majorana modes at both ends~\cite{Kitaev,Flensberg}.

Alternatively, we can define another pair of orthonormal states
\begin{eqnarray}
|L,\frac{\pi}{4}\rangle&=&|g\rangle+\tilde{c}_L^\dag|g\rangle, \\
|L,\frac{3\pi}{4}\rangle&=&|g\rangle-\tilde{c}_L^\dag|g\rangle.
\end{eqnarray}
Similarly, we have
\begin{eqnarray}
\gamma_{L,\frac{\pi}{4}}|L,\frac{\pi}{4}\rangle&=&|L,\frac{\pi}{4}\rangle\,,\\
\gamma_{L,\frac{3\pi}{4}}|L,\frac{3\pi}{4}\rangle&=&|L,\frac{3\pi}{4}\rangle\,.
\end{eqnarray}
Now we have a pair of Majorana modes are at site $L$, the right end of the chain. It can
be checked that no combination of $\ket{g}$ and $\tilde{c}_L^\dag|g\rangle$ can give us
a pair of Majorana modes, one at the left end and the other at the right end.

\subsection{Majorana modes in the bulk}
In the bulk we can  have a similar superposition state,
\begin{equation}
|j,\frac{\pi}{4}\rangle=|g\rangle+\tilde{c}_j^\dag|g\rangle, \quad \text {for} \ j\neq L\,.
\end{equation}
As one can readily check that
$\gamma_{j,\frac{\pi}{4}}|j,\frac{\pi}{4}\rangle=|j,\frac{\pi}{4}\rangle$, this is a
Majorana mode on bulk site $j$ of type $\frac{\pi}{4}$. However,   $|j,\frac{\pi}{4}\rangle$ is a
superposition of two different eigenstates of the Hamiltonian (\ref{Kitaev}) with the eigen-energy difference being $2\Delta$. This means that this Majorana mode will evolve dynamically.

Specifically, with the initial state $|\psi(0)\rangle=|j,\frac{\pi}{4}\rangle$, the system
evolves with time as (neglecting an overall phase),
\begin{equation}
|\psi(t)\rangle=|g\rangle+\tilde{c}_j^\dag|g\rangle e^{-i2\Delta t/\hbar}\,.
\end{equation}
The dynamics is cyclic with a  period of $T=\frac{\pi\hbar}{\Delta}$.
After a quarter of the period, $t=\frac{T}{4}=\frac{\pi\hbar}{4\Delta}$, the state evolves to
\begin{equation}
|\psi(\pi\hbar/4\Delta)\rangle=|j+1,\frac{\pi}{2}\rangle=|g\rangle-i\tilde{c}_j^\dag|g\rangle\,.
\end{equation}
This is a Majorana mode of type $\frac{\pi}{2}$ on site $j+1$.  After the second quarter period, the state becomes
\begin{equation}
|\psi(\pi\hbar/2\Delta)\rangle=|j,\frac{3\pi}{4}\rangle=|g\rangle-\tilde{c}_j^\dag|g\rangle.
\end{equation}
This is a Majorana mode of type ${\frac{3\pi}{4}}$ on site $j$.  After the third quarter period, the state becomes
\begin{equation}
|\psi(3\pi\hbar/4\Delta)\rangle=|j+1,0\rangle=|g\rangle+i\tilde{c}_j^\dag|g\rangle.
\end{equation}
This is a Majorana fermion of type ${0}$ on site $j+1$.  After a whole period of the evolution, the state returns to the initial state $|\psi(\pi\hbar/\Delta)\rangle=|j,\frac{\pi}{4}\rangle$. This
dynamical cycle can be summarized as
\begin{equation} \label{order}
|j,\frac{\pi}{4}\rangle\rightarrow|j+1,\frac{\pi}{2}\rangle\rightarrow|j,\frac{3\pi}{4}\rangle\rightarrow|j+1,0\rangle\rightarrow|j,\frac{\pi}{4}\rangle\,.
\end{equation}
In this sense, we can have Majorana modes in the bulk. The crucial difference
from the Majorana edge modes is that these bulk Majorana modes are not energy eigenstates
and oscillate dynamically.

Is it possible to have two Majorana modes simultaneously  along the Kitaev's chain? The answer is no.
As a simple example, we consider  the following  state,
\begin{equation}
|\psi(0)\rangle=|j_1,\frac{\pi}{4};j_2,\frac{\pi}{4}\rangle=(1+\tilde{c}_{j_1}^\dag)
(1+\tilde{c}_{j_2}^\dag)|g\rangle\,,
\end{equation}
where $j_1,j_2\neq L$. We have at site $j_1$
\begin{equation}
\gamma_{j_1,\frac{\pi}{4}}|j_1,\frac{\pi}{4};j_2,\frac{\pi}{4}\rangle=
|j_1,\frac{\pi}{4};j_2,\frac{\pi}{4}\rangle;
\end{equation}
However, at site $j_2$, we have
\begin{equation}
\gamma_{j_2,\frac{\pi}{4}}|j_1,\frac{\pi}{4};j_2,\frac{\pi}{4}\rangle=(1-\tilde{c}_{j_1}^\dag)(1+\tilde{c}_{j_2}^\dag)|g\rangle\neq|j_1,\frac{\pi}{4};j_2,\frac{\pi}{4}\rangle\,.
\end{equation}
This means that this state is a Majorana mode at site $j_1$ but not a Majorana mode at site $j_2$.
It appears that the existence of Majorana mode at site $j_1$ excludes  the emergence of Majorana fermion on site $j_2$.  According to the Hamiltonian (\ref{Kitaev}), this state  evolves with time as
\begin{equation}
|\psi(t)\rangle=(1+\tilde{c}_{j_1}^\dag e^{-i2\Delta t/\hbar})(1+\tilde{c}_{j_2}^\dag
e^{-i2\Delta t/\hbar})|g\rangle.
\end{equation}
It can  be shown easily that, at site $j_1$, the sequence in Eq. (\ref{order}) holds while
the dynamics at site $j_2$ does involve any Majorana mode.

In general we may construct the following state
\begin{eqnarray}
&&|j_1,\frac{\pi}{4};j_2,\frac{\pi}{4};\ldots;j_n,\frac{\pi}{4}\rangle\nonumber\\
&=&(1+\tilde{c}_{j_1}^\dag)(1+\tilde{c}_{j_2}^\dag)\ldots(1+\tilde{c}_{j_n}^\dag)|g\rangle\,,
\end{eqnarray}
where $j_1\neq j_2\neq\ldots\neq j_n$.
One can easily check that the Majorana sequence in Eq.(\ref{order}) occurs only at site $j_1$
and no Majorana modes at  any other sites. It is interesting to consider a special example
that is given by
\begin{eqnarray}
|\psi(0)\rangle&=&|L,\frac{\pi}{4};j_1,\frac{\pi}{4};\ldots;j_n,\frac{\pi}{4}\rangle\nonumber\\
&=&(1+\tilde{c}_{L}^\dag)(1+\tilde{c}_{j_1}^\dag)\ldots(1+\tilde{c}_{j_n}^\dag)|g\rangle\,.
\end{eqnarray}
It is clear that this state has a Majorana mode at site $L$.  Its dynamical evolution is
\begin{equation}
|\psi(t)\rangle=(1+\tilde{c}_{L}^\dag)(1+\tilde{c}_{j_1}^\dag e^{-i2\Delta t/\hbar})\ldots(1+\tilde{c}_{j_n}^\dag e^{-i2\Delta t/\hbar})|g\rangle.
\end{equation}
It can be checked easily that, at any time, the following relation holds
\begin{equation}
\gamma_{L,\frac{\pi}{4}}|\psi(t)\rangle=|\psi(t)\rangle\,.
\end{equation}
This shows explicitly that a Majorana mode of type $\frac{\pi}{4}$ is not affected by
the bustling dynamics at other sites, a feature that has inspired the current research
on Majorana physics in condensed matter systems.

One may have noticed an interesting feature in the above oscillating dynamics:
the dynamics is localized and the wave function of a Majorana fermion
can never spread to infinity. It is quite
peculiar as we know that the wave function of an electron in a real
vacuum always diffuses and can spread to infinity.
Localization in wave dynamics happens in rare occasions, such as Anderson localization in
random potentials and solitons in nonlinear media.

Our discussion so far  is done with the condition $\mu=0$, $t_p=d$.  When this condition
is slightly altered, the essential physics does not change.  The cyclic hopping can still occur.
The only difference  is that the wave function of the Majorana mode
spreads over several lattice sites, instead of the ideal localization
that we have with $\mu=0$, $t_p=d$.

The oscillatory process
is essentially a type of Zitterbewegung oscillation. To see this, we carry
out a Fourier transformation to the
momentum space, i.e., $c_k^\dag=\frac{1}{\sqrt{N}}\sum_jc_j^\dag \exp{({\rm i}jka)}$.
Without loss of generality, we assume $a=1$.
The Kitaev Hamiltonian in Eq.(\ref{Hamiltonian}) then becomes
\begin{equation}
\label{HamiltonianK}
H=\sum_k\left(\begin{array}{cc}c_k^\dag&c_{-k}\end{array} \right) H_k
 \left(\begin{array}{c}c_k\\c_{-k}^\dag\end{array} \right),
\end{equation}
where
\begin{equation} \label{effective}
H_{k}=\left(\begin{array}{cc}\xi(k)&\eta(k)\\ \eta(k)^*&-\xi(k)\end{array} \right).
\end{equation}
with $\xi(k)=-\mu-2t_p\cos(k)$ and $\eta(k)={\rm i}2\Delta\sin(k)$.  In the momentum space the Kitaev model is seen to assume a form identical to the BCS Hamiltonian.
If we regard particle and hole as two components of a pseudo-spin, then the rotation of
this pseudo-spin is governed by the  Hamiltonian $H_k$.
Due to the dependence of $H_{k}$ on $k$, the pseudo-spin
is coupled to momentum $k$. We have an effective spin-orbit coupling;
the cyclic dynamics in Eq.(\ref{order}) is essentially a type of
Zitterbewegung oscillation \cite{ZB}.

In summary, we have reviewed the concept of Majorana fermion as an elementary particle.
We pointed out that such a Majorana fermion does not exist in solid state systems
due to the lack of {\it real} anti-particle. However, despite this crucial difference,
one can regard hole as the anti-particle of electron in solid state physics and define
a Majorana operator. The physics related to this Majorana operator can be very interesting
and useful. We have illustrated it with  Kitaev's superconducting chain model.

\section{acknowledgement}

Q. Z. thanks Erhai Zhao for discussions on the Kitaev chain and the partial support by AFOSR FA9550-12-1-0079 as a visiting scholar at George Mason University.
This work is supported by the National Basic Research Program of China (Grants No. 2013CB921903) and the National Natural Science Foundation of China (Grants No. 11334001, and No. 11429402).

\end{document}